\documentclass[%
 reprint,
superscriptaddress,
 amsmath,amssymb,
 aps,
pra,
]{revtex4-1}

\usepackage{graphicx}
\usepackage{dcolumn}
\usepackage{bm}
\usepackage{hyperref}
\hypersetup{colorlinks = true, linkcolor = [rgb]{0.19411,0.51882,0.667058}, urlcolor = [rgb]{0.125490,0.29542,0.1647058}, citecolor = [rgb]{0.75882,0.37411,0.14117}}

\graphicspath{{../images/}}
\usepackage[toc,page]{appendix}

\newcommand{\ket}[1]{|{#1}\rangle}
\newcommand{\braket}[2]{\langle{#1}|{#2}\rangle}
\usepackage{braket}

\def\>{\rangle}
\def\<{\langle}

\usepackage[bottom]{footmisc}

\begin{document}

\title{Adaptive phase estimation with two-mode squeezed-vacuum and parity measurement}

\author{Zixin Huang}
\email[]{zixin.huang@sydney.edu.au}
\affiliation{School of Physics, University of Sydney, Sydney, NSW 2006, Australia}

\author{Keith R. Motes }

\affiliation{Department of Physics and Astronomy, Macquarie University, Sydney NSW 2113, Australia}

\author{Petr M. Anisimov}

\affiliation{Los Alamos National Laboratory, New Mexico 87545}

\author{Jonathan P. Dowling}
\affiliation{Hearne Institute for Theoretical Physics and Department of Physics \& Astronomy, Louisiana State University, Baton Rouge, LA 70803}

\author{Dominic W. Berry}
\affiliation{Department of Physics and Astronomy, Macquarie University, Sydney NSW 2113, Australia}

\begin{abstract}
 A proposed phase-estimation protocol based on measuring the parity of a two-mode squeezed-vacuum state at the output of a Mach-Zehnder interferometer shows that the Cram\'{e}r-Rao sensitivity is sub-Heisenberg [Phys.\ Rev.\ Lett.\ {\bf104}, 103602 (2010)]. 
However, these measurements are problematic, making it unclear if this sensitivity can be obtained with a finite number of measurements. This sensitivity is only for phase near zero, and in this region there is a problem with ambiguity because measurements cannot distinguish the sign of the phase. Here, we consider a finite number of parity measurements, and show that an adaptive technique gives a highly accurate phase estimate regardless of the phase. We show that the Heisenberg limit is reachable, where the number of trials needed for mean photon number $\bar{n}=1$ is approximately one hundred. We show that the Cram\'{e}r-Rao sensitivity can be achieved approximately, and the estimation is unambiguous in the interval ($-\pi/2, \pi/2$).
\end{abstract}

\date{\today}

\maketitle

\thispagestyle{empty}

\section{Introduction}

Phase estimation and optical interferometry are the basis for many precision measurement applications. Coherent light based interferometry is most commonly used but its sensitivity for phase estimation is limited by the shot-noise limit, \mbox{$(\Delta \varphi)^{2}\ge \bar{n}^{-1}$}, where $\varphi$ is the unknown phase, and $\bar{n}$ is the mean number of photons used to perform the estimation \cite{Dowling2008}. This is not a problem in the case of limitless resources or in the case of samples that can withstand large doses of radiation. However, in order to achieve a finer precision given a finite amount of resources, one has to resort to interferometry with quantum states of light, such as N00N states \cite{LeeDowling2002} with parity measurements \cite{a:Gerry.2010, PhysRevA.68.023810},
in order to achieve sub-shot-noise or the Heisenberg limit (HL) to sensitivity of phase estimation. Squeezed vacuum, which is the brightest experimentally available nonclassical light, has received much attention. In particular, the two-mode squeezed-vacuum (TMSV) which is the simplest two-mode state that contains strong photon-number entanglement, offers a notable improvement in phase estimation precision when compared to coherent states \cite{PhysRevLett.104.103602,gao2010super,PhysRevA.68.023810,seshadreesan2011parity}.

\begin{figure}[t]
\includegraphics[trim= 0 1.5cm 0 0, clip,width=3.0in]{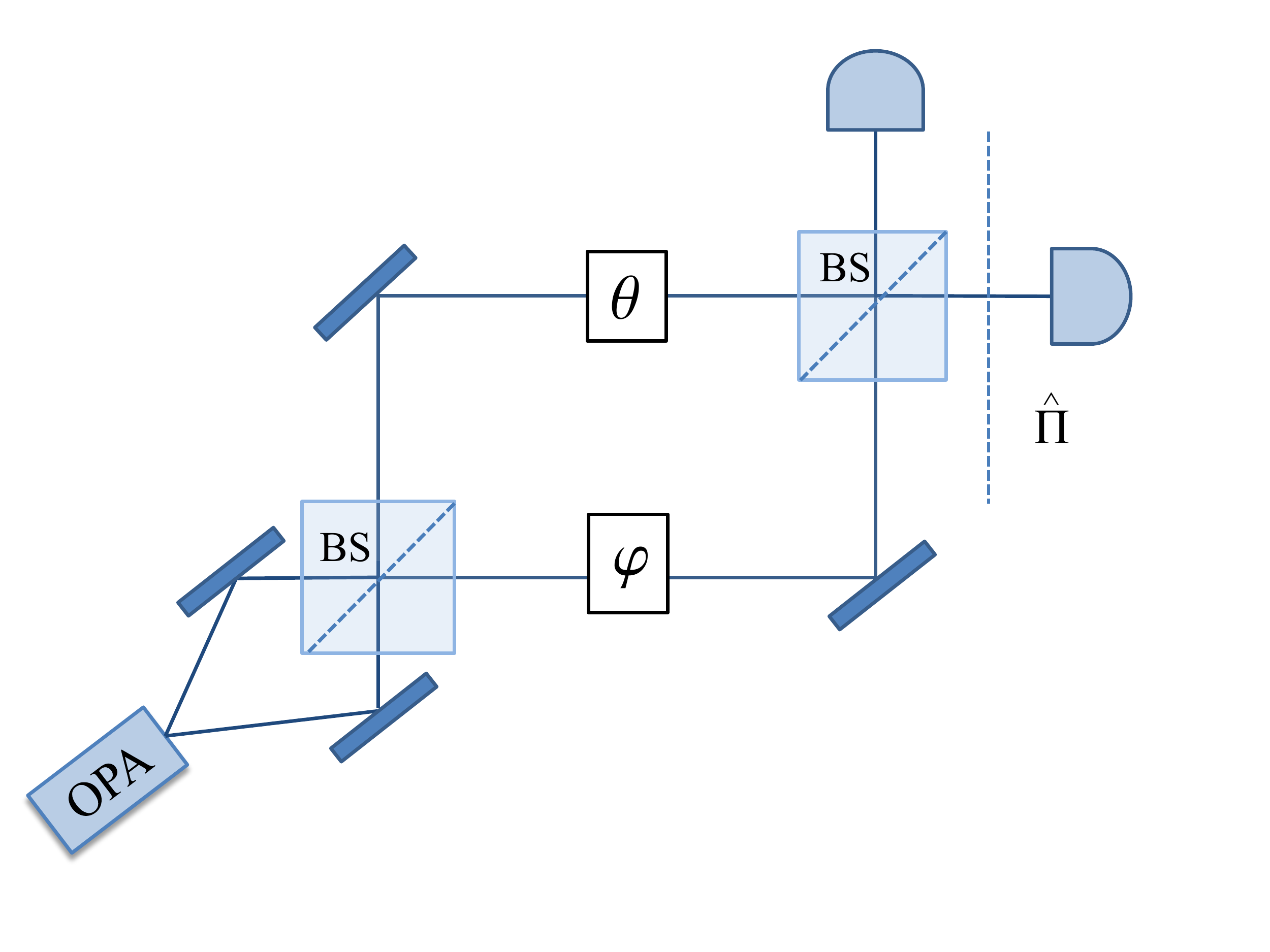}
\caption{\label{fig:MZI} Two-mode squeezed-vacuum states are generated at the input of the Mach-Zehnder interferometer (MZI) by an optical parametric amplifier (OPA), where $\varphi$ is the phase being measured and $\theta$ is a controllable phase. The parity signal at the output of the MZI is then measured with photon-number-resolving detectors. Since TMSV states always have even photon numbers, the parity signal can be detected by performing photon-counting at only one output.
}
\end{figure}

Significant advances have been made in quantum-enhanced phase
sensitivity \cite{a:Lloyd.2011} and the meaning of the Heisenberg limit has been
thoroughly examined \cite{bib:zwierz2012ultimate, bib:PhysRevA.85.041802}. 
Yet, a proposed phase estimation scheme dips below the HL in the case of an infinite number of parity
measurements \cite{PhysRevLett.104.103602}. 
That scheme is based on measuring the parity of the state of light at the output of a Mach-Zehnder
interferometer (MZI), as shown in Fig.~\ref{fig:MZI}, with two-mode squeezed-vacuum input.
A parity measurement focuses on whether the output photon number is odd or even, rather than the 
the actual number itself. 
 It turned out that this particular scheme using TMSV input has sub-Heisenberg sensitivity, due to
the fact that the photon number uncertainty for the state of light inside of the
MZI is greater than the average photon number used for the measurement
\cite{PhysRevA.79.033822,PhysRevLett.105.120501}.

The Heisenberg limit, $\Delta\varphi^2 = 1/(M N^2)$ for states with $N$ fixed total number of photons such as twin-Fock \cite{bib:holland1993interferometric} or N00N states \cite{LeeDowling2002}
and $M$ copies of the state, is a rigorous lower limit for \textit{local} phase sensitivity for such states \cite{PhysRevLett.99.070801}. However, for states with well-defined mean photon number but undefined total photon number, such as the TMSV used here, it is now understood that the Heisenberg limit so defined is not a lower limit.

Here we concern ourselves primarily with the Cram\'{e}r-Rao bound, which is provably the ultimate limit of phase sensitivity \cite{bib:demkowicz15}. That is, for a measurement providing an unbiased estimate of a parameter $\varphi$, the variance of the estimate can be no less than the Cram\'{e}r-Rao bound. Hence, if we saturate this limit, as we do in this work with parity detection, then our measurement scheme is optimal. 

Unlike in Ref.~\cite{PhysRevLett.104.103602} where the Cram\'{e}r-Rao bound sensitivity is expected in the limit of an infinite number of parity measurements, we consider a finite parity measurement record. We use the terminology ``detection" to denote measurement on an individual copy of the state, and ``measurement record" to denote the list of parity detections.

\begin{figure}[tb]
\includegraphics[width=3.1in]{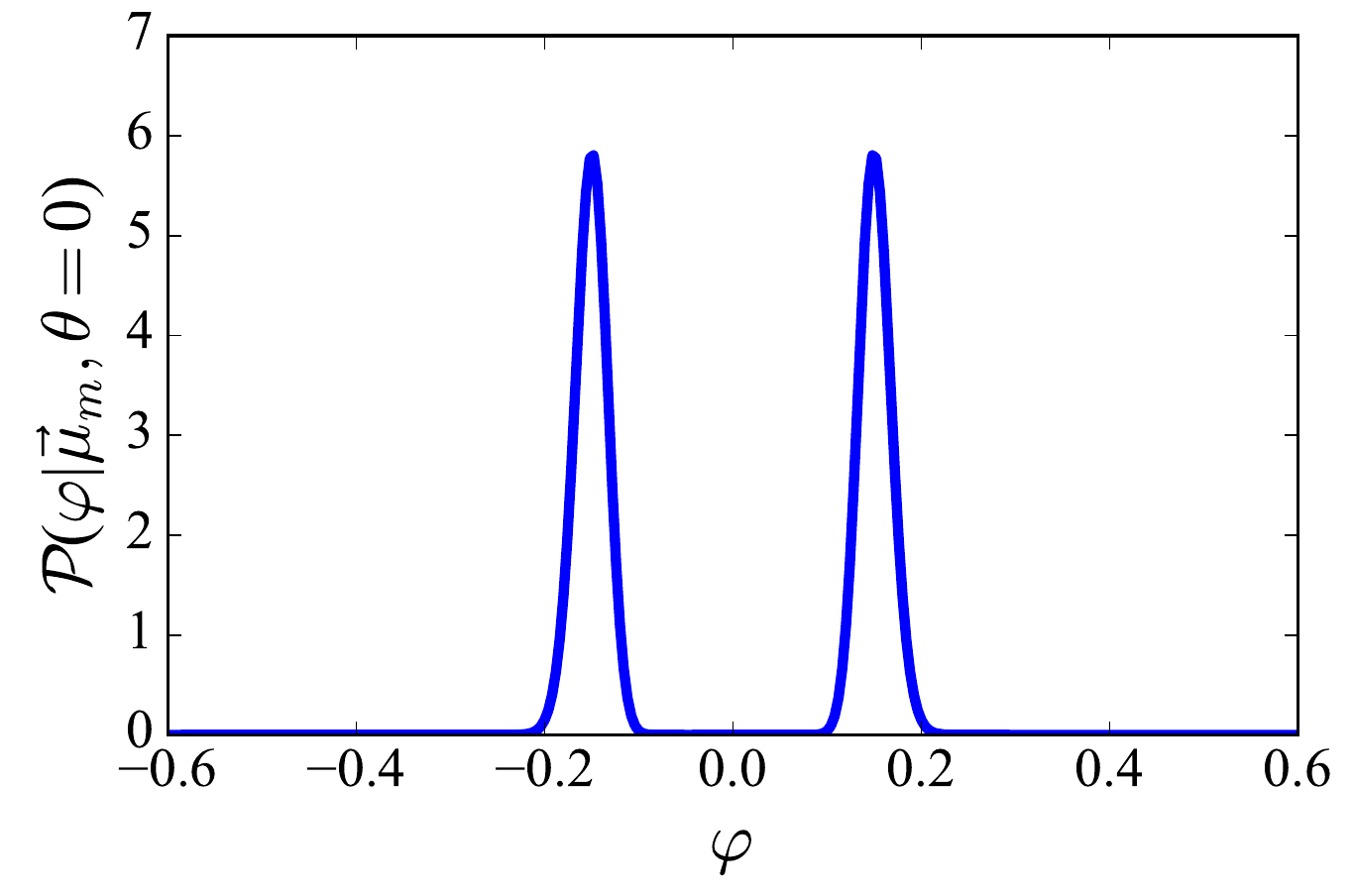} 
\caption{The probability distribution for the phase $\varphi$ for an example measurement record, generated for an actual phase of $\varphi = 0.15$ and the control phase $\theta = 0$. The record consists of $M =512$ parity detections, of which 466 are even.}
\label{fig:amb}
\end{figure}

In previous schemes considered \cite{gao2010super,PhysRevA.68.023810,seshadreesan2011parity}, it was unclear as to how an unambiguous estimate could be made, given that the parity of the output is symmetric around the origin. If one infers the value of the measured phase $\varphi$ by only 		considering the statistics (number of odd/even outcomes) of a static interferometer, there is ambiguity in the sign of the phase estimate.
Figure~\ref{fig:amb} shows the probability distribution for the phase $\varphi$ for an example measurement record, generated for an actual phase of $0.15$. 
The distribution has two maxima distributed symmetrically around $ 0$. 

In addition, the most sensitive region is confined to a small interval where the relative phase between the two arms of the interferometer is zero, where the ambiguity of the sign is most problematic. Moving away from this region, the phase sensitivity decays quickly.

\begin{figure}[tb]
\includegraphics[width=3.1in]{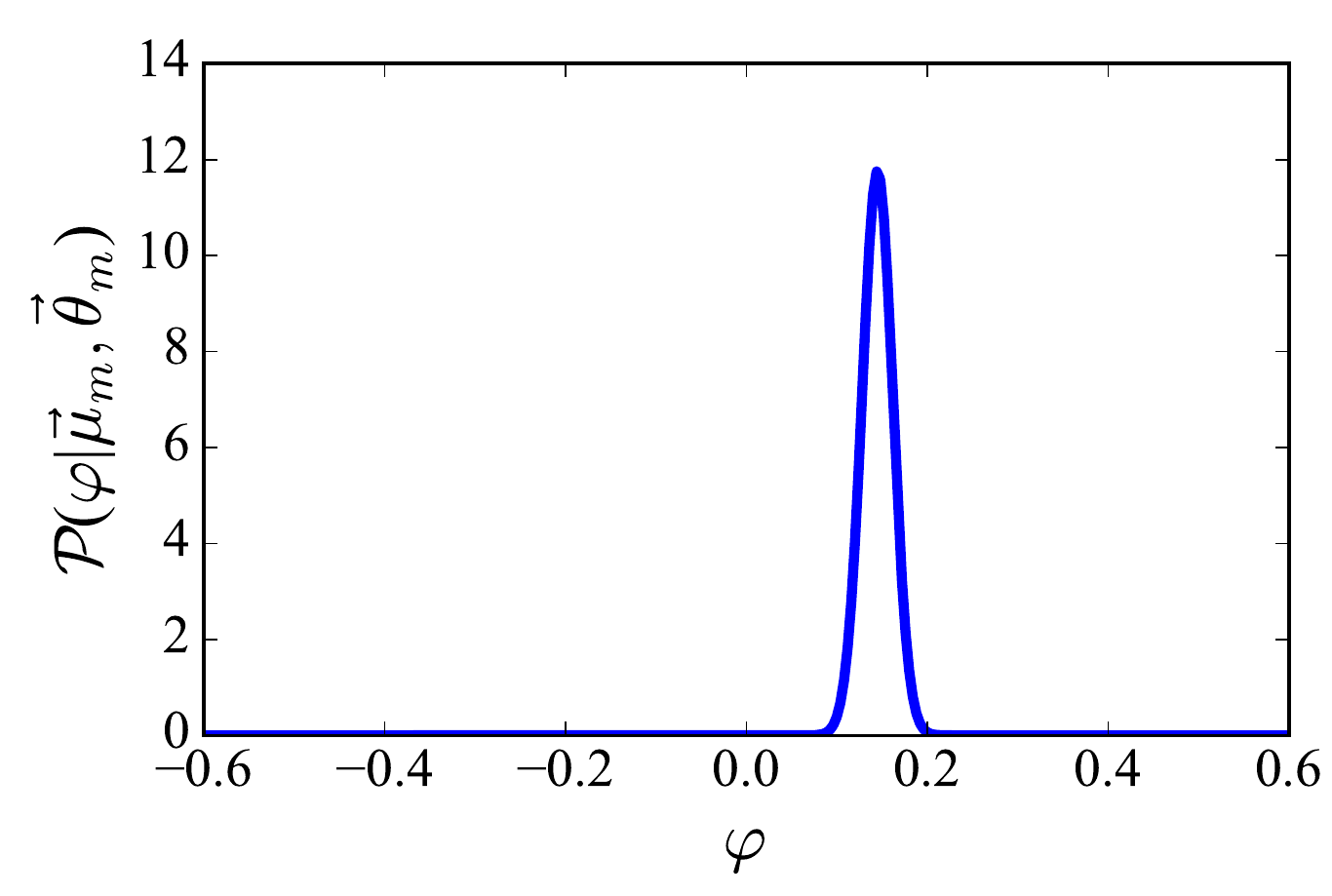}
\caption{The probability distribution for the phase $\varphi$ for an example measurement record, generated for an actual phase of \mbox{$\varphi = 0.15$} and $\theta$ being adaptive. The record consists of $M =512$ parity detections. There is no ambiguity in the sign of $\varphi$.}
 \label{fig:unamb}
\end{figure}

We introduce a feedback technique described in the following sections and eliminate this ambiguity; an example of the probability distribution for the phase is shown in Fig.~\ref{fig:unamb}. By implementing an adaptable control phase $\theta$ we extend the region where the measurement is accurate to the interval ($-\pi/2, \pi/2$).
We show that with a sufficient number of detections, the method achieves Heisenberg-limited sensitivity.
In this paper we define the Heisenberg limit as $\Delta\varphi^2 \equiv 1/ (M\bar{n}^2)$
 \cite{Dowling2008}.

The structure of the paper is as follows. In Sec.~\ref{sec:2} we describe the model of the TMSV state and discuss the problematic features associated with parity measurements. In Sec.~\ref{sec:3} we detail the adaptive technique used to resolve these issues, and present results for the accuracy of the measurement scheme. We discuss the effects of photon loss in Sec.~\ref{sec:4}, then conclude in Sec.~\ref{conc}.

\section{Model \label{sec:2}}
We consider a phase estimation scheme with a two-mode squeezed-vacuum input state which is commonly generated in unseeded optical parametric amplifiers. A TMSV state is ideally a superposition of twin Fock states,
\begin{align}
\ket{ \psi_{\bar{n} }}=\sum \limits_{n=0}^{\infty} \sqrt{p_n\!\left(\bar{n}\right)}\ket{ n,n},
\label{eq:tvsm}
\end{align}
where $p_n\!\left(\bar{n}\right)=(1-t_{\bar{n}})t^n_{\bar{n}}$,
$t_{\bar{n}}=1/\left(1+2/\bar{n}\right)$, and $\bar{n}$ is the average photon number in the state \cite{gerry2005introductory}.

Parity based phase estimation was originally introduced in quantum optics by Gerry \cite{PhysRevA.61.043811} and is based on the parity of the photon number detected at the output of the MZI.
Since the photon number in a TMSV state is always even, the parity signals are the same in both the output ports, and therefore can be detected by performing photon-counting at only one port.
It turns out that parity detection is sufficient to achieve the Cram\'er-Rao bound in the case where the state is path symmetric \cite{seshadreesan2013}, which is the case here.
Propagation of the light through a MZI imprints phase information on the state that is retrieved by measuring parity at the output of the MZI. The expected value of the
parity signal $\langle\hat{\Pi}\rangle$ for a TMSV based phase estimation scheme, 
\begin{align}
\langle\hat{\Pi}\rangle=\frac{1}{\sqrt{1+\bar{n}(\bar{n}+2)\sin^{2}(\theta-\varphi)}},
\label{eq: Parity}
\end{align} 
was obtained in Ref.~\cite{PhysRevLett.104.103602}. Equation~\eqref{eq: Parity} was obtained by summing the expected parity value $\braket{\hat{\Pi}}_n$ for each twin-Fock state $\ket{n,n}$, weighted by their respective 
probability of occurring,
\begin{align}
\braket{\hat{\Pi}} = (1-t_{\bar{n}}) \sum^{\infty}_{n=0} t_{\bar{n}} ^n \braket{\hat{\Pi}}_n,
\label{eq:sumparity}
\end{align}
where $\braket{\hat{\Pi}}_n = (-1)^n \mathcal{L}_n[\cos(2(\theta - \varphi+\frac{\pi}{2}))]$ and $\mathcal{L}_n$ is the Legendre polynomial  of order $n$ \cite{PhysRevA.68.023810}.

A straightforward method for determining the magnitude of the unknown phase $\varphi$ is as follows. If we let the controllable phase be $\theta=0$ and we know $\bar{n}$ at the input, then we send the TMSV through the MZI of Fig.~\ref{fig:MZI}, which interrogates the unknown phase $\varphi$. Then, we perform parity measurements at the output, which returns either an even or odd outcome with probabilities $\mathcal{P}_{\rm e}$ and $\mathcal{P}_{\rm o}$ (as defined below), respectively. This allows us to determine the parity signal $\hat{\Pi}$. 

This parity measurement can be implemented with photon-number-resolving detectors or homodyne detection \cite{1367-2630-13-8-083026,1367-2630-12-11-113025}. 
Inferring the parity of a state disregards the actual number of photons detected and focuses on
whether this number is even or odd. Since $\mathcal{P}_{e}+\mathcal{P}_{o}=1$ and the expectation value of a state's parity is $\langle\hat{\Pi}\rangle=\mathcal{P}_{e}-\mathcal{P}_{o}$, the probabilities of detecting an even or odd photon number are
\begin{align}
\mathcal{P}_{\rm e} &= \frac{1}{2}(1+ \braket{\hat{\Pi}}), \qquad
\mathcal{P}_{\rm o} = \frac{1}{2}(1-\braket{\hat{\Pi}}).
  \label{eq:probs}
\end{align}
When the estimate is unbiased \cite{bib:PhysRevA.86.053813}, the precision of the estimate is lower-bounded by the Cram\'{e}r-Rao bound,
$\Delta \varphi^2 \leq 1/ M \mathcal{F}(\varphi)$, where $M$ is the number of times the estimation is repeated and $\mathcal{F}(\varphi)$ is the 
Fisher information \cite{bib:demkowicz15}. 
In this case, the Fisher information is 
\begin{align} 
\mathcal{F}(\varphi) &= \frac{1}{\mathcal{P}_{\rm e}}\left(\frac{\partial\mathcal{P}_{\rm e}}{\partial\varphi}\right)^2 + \frac{1}{\mathcal{P}_{\rm o}}\left(\frac{\partial\mathcal{P}_{\rm o}}{\partial\varphi}\right)^2 \nonumber \\
&= \frac 1{1-\braket{\hat{\Pi}}^2}\left( \frac{\partial\braket{\hat{\Pi}}}{\partial\varphi} \right)^2 \nonumber \\
&=\frac{\cos^2(\theta-\varphi) \bar{n} (\bar{n}+2)}{ \left[1+ \bar{n}(\bar{n}+2) \sin^2(\theta-\varphi)\right]^2}.
 \label{eq:crb}
\end{align}
 The Fisher information is maximized for \mbox{$\theta-\varphi =0$}. Therefore, the ultimate precision of this estimation scheme is \mbox{$1/M \bar{n}(\bar{n}+2)$}, which is sub-Heisenberg if the bound can be achieved. When $F \geq \bar{n}$, the scheme performs better than the shot-noise limit. This would be achieved if $|\varphi-\theta|$ is smaller than approximately $\bar{n}^{-1/4}$. As $\varphi$ increases, the sensitivity decays quickly.

One can estimate $\varphi$ from the statistics of the detections.  In order to choose the estimate, it is useful to determine a probability density function for the phase based on the detection results.  This probability density can be determined via Bayes' theorem to be
\begin{align}
\mathcal{P}(\varphi|\ell) \propto \mathcal{P}_e^\ell (\varphi) \mathcal{P}_o ^{M-\ell}(\varphi),
\label{eq:eo}
\end{align}
where $M$ is the number of parity detections and $\ell$ is the number of even results.  The estimate of $\varphi$ could be chosen to be, for example, the maximum of this probability distribution.

One can model phase estimation with TMSV and parity detection numerically in the following way. Choose an average photon number $\bar n$ and an unknown phase $\varphi$, then numerically generate a measurement record of finite length $M$ using the probabilities ${\cal P}_e$ and ${\cal P}_o$. From this measurement record the probability density function for $\varphi$ can be calculated using Eq.~\eqref{eq:eo}. An example of a simulation is presented in Fig.~\ref{fig:amb} for the case where $\theta$ is constant. With $\theta$ constant, there is ambiguity in the estimation of $\varphi$, because the distribution is symmetric about 0.

\section{The Measurement scheme \label{sec:3}}

In order to eliminate this ambiguity, we apply a feedback technique, which changes the controlled phase $\theta$ based upon previous detection results and controlled phases. Intuitively, this method works by maximizing \mbox{$|\braket{e^{i \varphi}}|$} thereby reducing the appearance of multiple peaks \cite{PhysRevLett.85.5098}. Since,
\begin{align}
\braket{e^{i\varphi}} = \int \limits_{0}^{2\pi} e^{i \varphi} \mathcal{P}(\varphi)   d\varphi  ,
\end{align}
\mbox{$|\braket{e^{i \varphi}}|$} is maximum when $\mathcal{P}(\varphi)$ is a delta function, and equals zero when $\mathcal{P}(\varphi)$ is flat. Here we use a variation of the method in Ref.~\cite{PhysRevLett.85.5098} where we maximize $|\braket{e^{i 2\varphi}}|$, as explained next.

The adaptive technique uses the latest probability distribution $\mathcal{P}(\varphi)$ to calculate the next controlled phase.
Initially the distribution is flat, because there is no phase information.
The initial controlled phase is therefore chosen to be random.
Then, after each detection, the probability distribution is updated using Bayes' theorem,
\begin{align}
\mathcal{P}(\varphi|\vec\mu_m, \vec\theta_m ) \propto \mathcal{P}(\mu|\varphi, \vec\theta_m ) \mathcal{P}(\varphi| \vec\mu_{m-1}, \vec\theta_{m-1} ),
\label{eq:bayes}
\end{align}
where $\mu$ denotes whether the detection is even or odd, $\vec\mu_m = (\mu_1, \mu_2,.., \mu_m)$ is the vector of successive detection results, and $\vec\theta_m =(\theta_1, \theta_2,.., \theta_m)$ is the vector of the corresponding controlled phases.
Since the probabilities are periodic with period $\pi$, they can be expressed as a Fourier series,
\begin{align}
\mathcal{P}(\varphi|\vec\mu_m, \vec\theta_m ) = \frac{1}{2\pi} \sum\limits_{j=-x}^x a_j e^{2 i j \varphi},
\label{eq:fourier}
\end{align} 
where $a_j$ is complex, and depends on $\vec\mu_m$ and $\vec\theta_m$.

Before the first detection, Eq.~\eqref{eq:fourier} contains only one term, $a_0 = 1/2\pi$. After each detection result given by the probabilities in Eq.~\eqref{eq:probs}, the Fourier coefficients $a_j$ in Eq.~\eqref{eq:fourier} are updated using Eq.~\eqref{eq:bayes}, which again uses Eq.~\eqref{eq:probs}.
However, to exactly represent Eq.~\eqref{eq:probs} in terms of Fourier coefficients, an infinite sum over the Legendre polynomial terms is needed in Eq.~\eqref{eq:sumparity}. In order to perform the numerical calculations, we truncated this infinite sum.
It was found that the cut-off needed depended on $\bar{n}$; the values used are given in Table \ref{table:one}.

\begin{table}[tb]
  \begin{tabular}{ | c | c | c| c|} 
    \hline
    $\bar{n}$ 	&No. of terms 
    \\ \hline
     1    		&  10 									 \\ 
     2   		&  10 									\\
     3  		&  15 								 \\
     
     5			&  20  							 \\
   
          8    		&  25  							\\
    \hline
  \end{tabular}
  \caption{\label{table:one} For a TMSV state of mean photon $\bar{n}$, the cut-off for the number of Legendre polynomial terms used in Eq.~\eqref{eq:sumparity}.}
\end{table}

We adjust the controlled phase $\theta_m$ based on the previous detection results and controlled phases. The value for $\theta_m$ is the one which maximizes the average sharpness of the probability distribution for $\varphi$ after the next detection, across the interval ($-\pi/2,\pi/2$).
Here we take the sharpness to be 
$s(\theta)\equiv|\braket{e^{i 2\varphi}}|$, which differs from the case in Ref.~\cite{PhysRevLett.85.5098} where the sharpness is $|\braket{e^{i\varphi}}|$.
We make this choice because the probability distribution for the parity detection results has a period of $\pi$ instead of $2\pi$.

The explicit expression for the average sharpness is
\begin{align}
s_{\text{av}}(\theta_m)= \frac{1}{2\pi} \sum\limits_{\mu= \{+1, -1\} } \left|\int \limits_{0}^{\pi} e^{i 2\varphi}  \prod \limits_{k=1}^{m} \mathcal{P}(\mu_k|\varphi, \vec\theta_k)   \mathrm{d} \varphi \right|.
\label{eq:sharpness}
\end{align}
This expression corresponds to the average sharpness for the measurement results weighted by their probability of occurring.
Maximizing this expression yields the highest average accuracy of the phase estimates after the next detection \cite{PhysRevA.80.052114}.
The integral over $\varphi$ simply yields the coefficient $a_{-2}$. Therefore, this integral may be obtained by summing the absolute value of $a_{-2}$ for \mbox{$\mathcal{P}(\varphi|\vec\theta_m,\vec\mu_{m-1},\mu_m= \text{even})$} and \mbox{$\mathcal{P}(\varphi|\vec\theta_m,\vec\mu_{m-1},\mu_m = \text{odd})$} after the next detection. 
No analytical formula exists for calculating the optimal $\theta_m$, so it was determined numerically \cite{PhysRevA.90.023856}. 
%
%

At the end of each measurement record, the estimate is taken to be the argument of $a_{-2}$, corresponding to $\arg(\braket{e^{2 i \varphi}})$.
This estimate of the phase is optimal for a measure of the measurement accuracy based on $|\braket{\cos(2 (\hat \varphi-\varphi)}|$ \cite{PhysRevA.80.052114}, where $\hat\varphi$ is the value of an individual estimate.
Here we are using the mean-square error (MSE), for which this estimate is not exactly optimal.
(The MSE is an estimate of the error that is equal to the variance if the measurement is unbiased, and also appropriately penalises biased estimates.)
However, this estimate is close to optimal for narrowly peaked distributions.
In that case, the cosine function can be accurately approximated by expanding to second order, so
\begin{equation}
|\braket{\cos(2 (\hat \varphi-\varphi)}| \approx 1-2\braket{(\hat \varphi-\varphi)^2}.
\end{equation}
Therefore the estimate that maximizes $|\braket{\cos(2 (\hat \varphi-\varphi)}|$ approximately minimizes the MSE
$\braket{(\hat \varphi-\varphi)^2}$.
This estimate of the phase is also close to unbiased for narrowly peaked distributions.
This is because $\braket{e^{2 i (\hat \varphi-\varphi)}}=1$.
For narrowly peaked distributions, the exponential can be expanded to first order, giving $1+\braket{2 i (\hat \varphi-\varphi)}\approx 1$, which implies that $\braket{\hat \varphi}\approx \varphi$.
Therefore, the Cram\'er-Rao bound should hold approximately.

Another feature of this estimate of the phase, is that together with the random initial controlled phase, it ensures that the measurement scheme is covariant.
That is, the probability distribution for the error in the estimate is independent of the system phase.
In order to fairly evaluate the overall performance of a measurement scheme, one should determine the MSE averaged over the system phase.
When the measurement scheme is covariant, this averaging is unnecessary, because the MSE is independent of the system phase.

In order to estimate the performance of this adaptive scheme, measurement records were generated numerically and the controlled phases $\theta_m$ were calculated.  For a measurement record consisting of $M$ detections, we generated $J$ records to evaluate the MSE of the estimates. Since it becomes computationally intensive to obtain a large number of estimates as $M$ increases, for each different $M$, $J$ was chosen such that the uncertainty in the estimated MSEs is less than $ 3\%$ \cite{Rose2002}. It was found that in order to achieve the same precision in calculating $\Delta\varphi^2$, as $M$ increases, $J$ can be reduced. For example, for $M \in [64,128]$, $J =10^6$ phase estimates were performed to calculate $\Delta\varphi^2$, whereas when $M = 3096$, only $J=5000$ estimates were necessary.

\begin{figure}[tb]
\includegraphics[ width=3.1in]{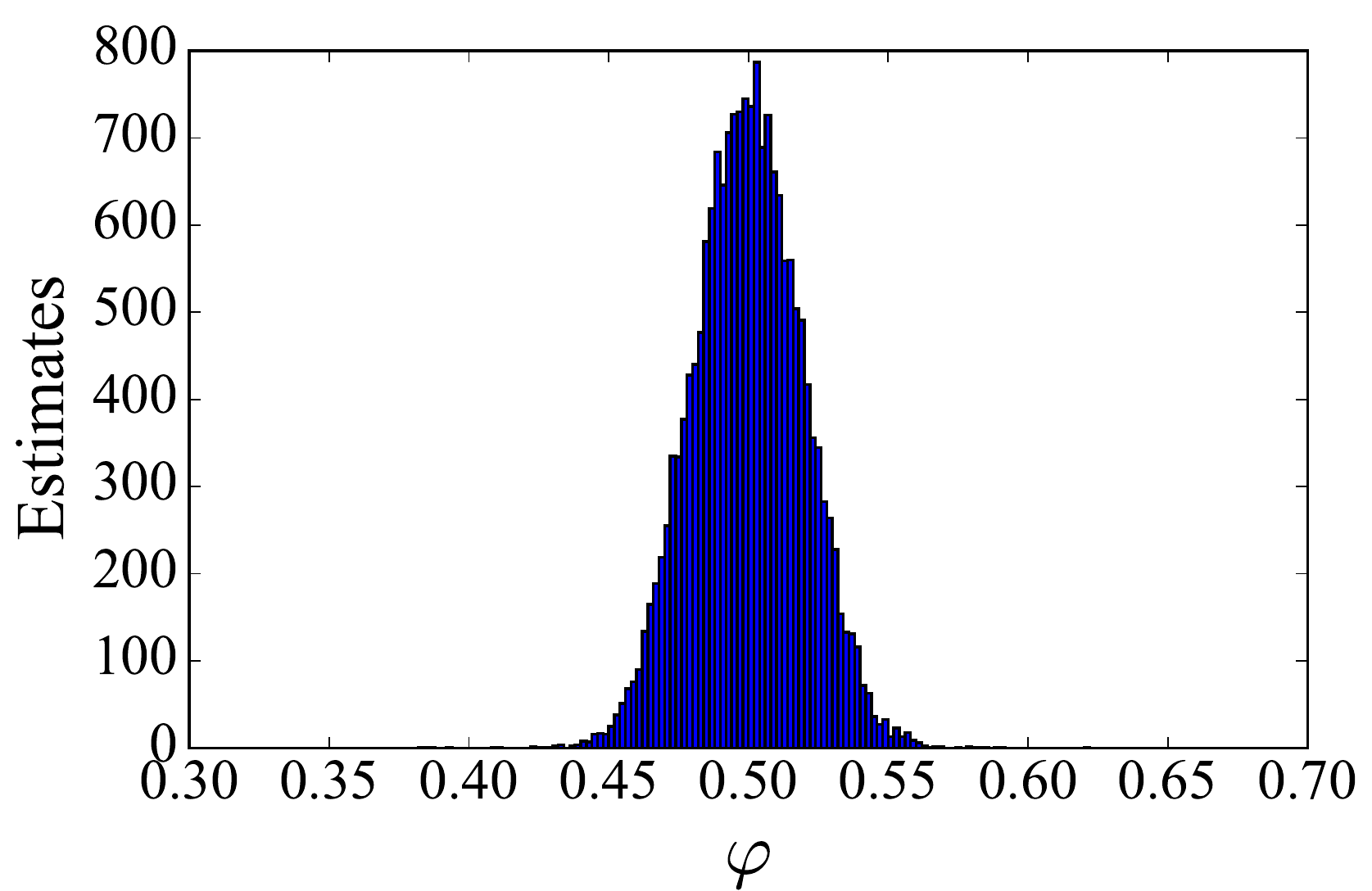}
\caption{The distribution of estimates for $\varphi$, where $M=256$ detections were used to obtain each estimate. The plot contains results from $J = 20000$ measurement records which were numerically generated for $\varphi = 0.5$ and $\bar{n}=3$. The distribution has a MSE $\Delta\varphi^2=3.81 \times 10^{-4}$. }
\label{fig:distribution}
\end{figure}

As an example, $J =20000$ measurement records were generated for $\varphi=0.5$ and $\bar{n}=3$, where $M =256$ detections were used for each estimate. Figure~\ref{fig:distribution} shows the distribution of the estimates. There is a spread in the phase estimates with a MSE $\Delta \varphi^2=3.81 \times 10^{-4}$. 

\begin{figure}[tbh]
\includegraphics[width=0.49\textwidth]{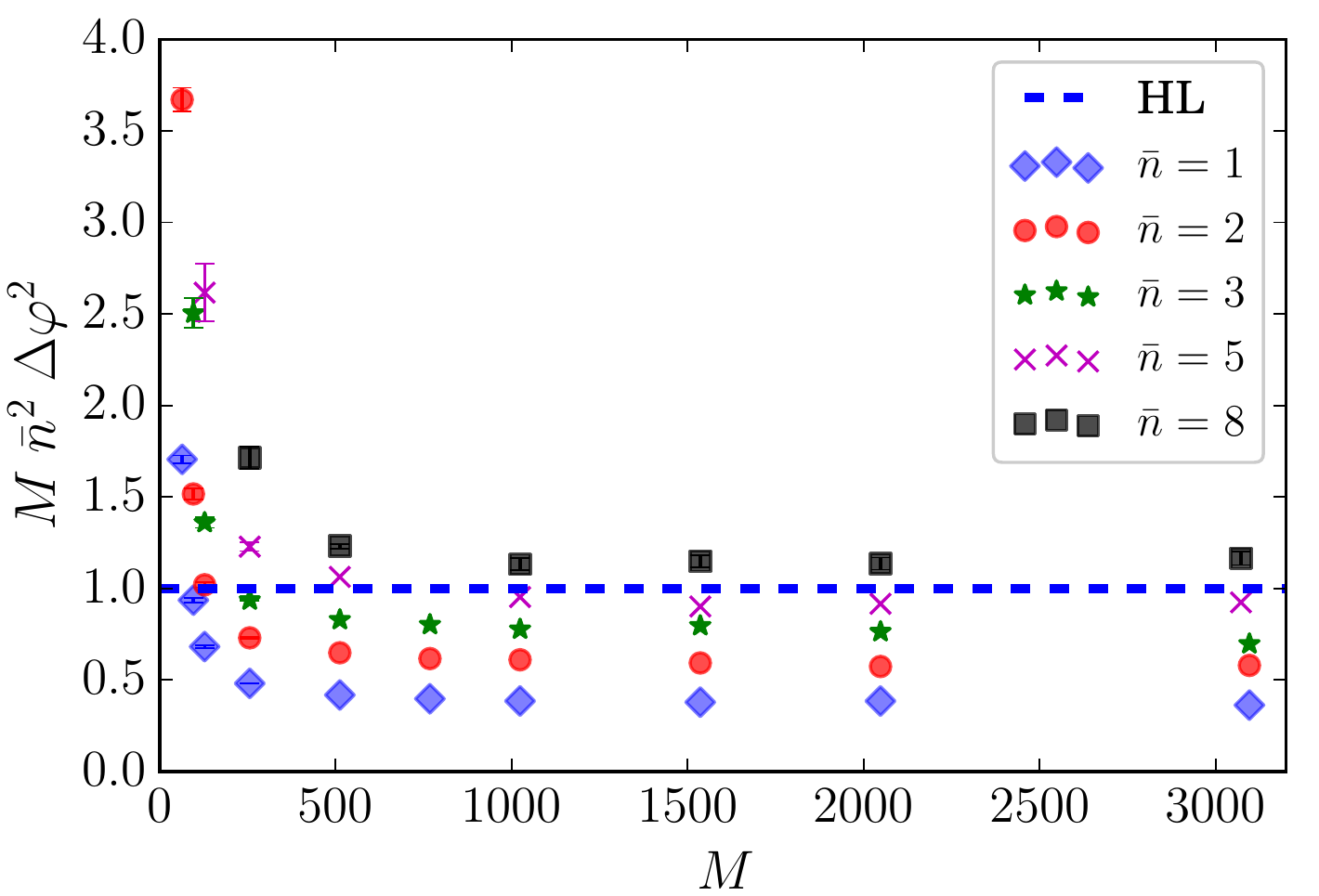} 
\caption{The ratio of the phase MSE to the Heisenberg limit versus the measurement record length $M$, for TMSV states with a range of mean photon numbers: $\bar{n} = 1$ (blue diamonds), $\bar{n}=2$ (red circles), $\bar{n}=3$ (green stars), $\bar{n}=5$ (purple crosses) and $\bar{n}= 8$ (black squares). Error bars are shown only if they are larger than the marker size. The Heisenberg limit (blue dashed line) is plotted for comparison (it is $1$ because all values are shown as a ratio to the Heisenberg limit). 
}
\label{fig:allvariances}
\end{figure}

\begin{figure}[tbh]
\includegraphics[width=0.49\textwidth]{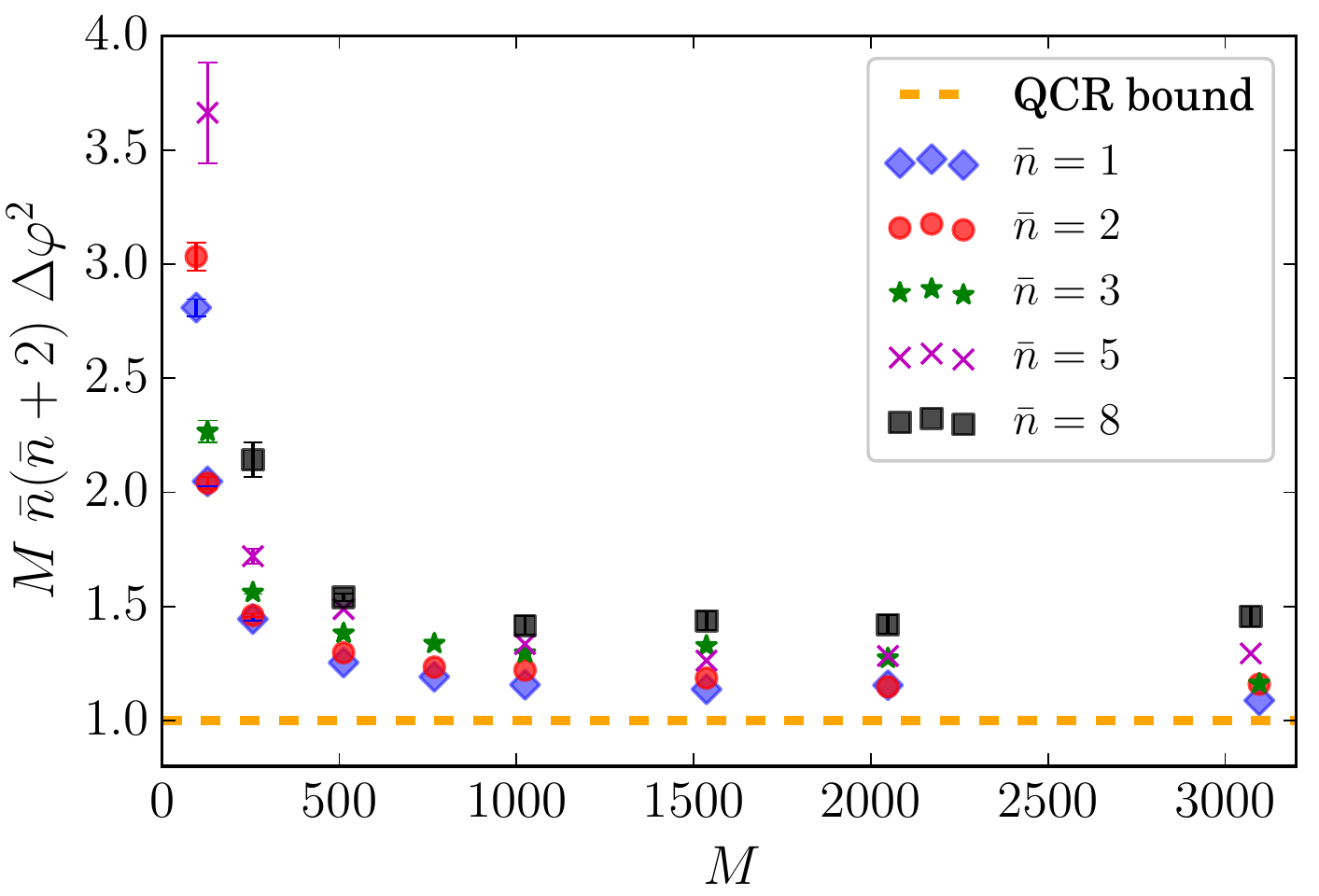} 
\caption{The ratio of the phase MSE to the quantum Cram\'{e}r-Rao bound versus the measurement record length $M$, for TMSV states with a range of mean photon numbers: $\bar{n} = 1$ (blue diamonds), $\bar{n}=2$ (red circles), $\bar{n}=3$ (green stars), $\bar{n}=5$ (purple crosses) and $\bar{n}= 8$ (black squares). The quantum Cram\'{e}r-Rao bound (dashed orange line) is shown for comparison. The error bars are shown only if they are larger than the size of the markers.
}
\label{fig:QCR}
\end{figure}

In Fig.~\ref{fig:allvariances} we show the ratio of the MSE to the Heisenberg limit against the length of the measurement record $M$. For most of the data points, the error bar is smaller than the marker. For $1.0 \leq \bar{n} \leq 5.0$, the MSEs beat the HL. (Since the HL is $\Delta \varphi^2 = 1/(\bar{n}^2 M)$, when multiplied by $M\bar{n}^2$, the HL equals 1 on this plot.)
In Fig.~\ref{fig:QCR}, we show the ratio of the MSE to the quantum Cram\'{e}r-Rao bound; as we expect, the MSEs asymptotically approach this limit. Evidently, the smaller the mean photon number, the faster the MSEs converge to the bound. For $\bar{n} = 1$, when $M=3096$, the phase MSE is larger than the Cram\'{e}r-Rao bound by less than $10\%$. For larger $\bar{n}$, the relative difference from the Cram\'{e}r-Rao bound is larger for the same $M$. Larger $M$ would be needed to achieve agreement within 10\%.

\section{Measurement scheme with losses \label{sec:4}}

In reality photon-number-resolving detectors are not 100$\%$ efficient, and photonic states in an interferometer are subjected to  loss. If losses are equal in both arms, then the inefficiency of the system can be combined and described by a single parameter $\eta$, where $1-\eta$ is the probability of losing a photon.
To model detector inefficiency we use \cite{bib:motes2013spontaneous},

\begin{align} 
\mathcal{P}_\mathrm{D}(t|s) = {s \choose t} \eta^t (1-\eta)^{s-t},
\label{eq:detectorloss}
\end{align}

\noindent which gives the conditional probability of detecting $t$ photons given that $s$ photons were present.
We combine Eqs.~\eqref{eq:tvsm} and \eqref{eq:detectorloss} to determine the Fourier coefficients for the parity signal in the presence of loss, which are then used to perform the simulations to calculate the phase MSE. When $\eta$ is less than $1$, the visibilities of the signal peaks as a function of $\varphi$ are reduced; this effect becomes more pronounced as $\bar{n}$ is increased.


\begin{figure}[tb]
\includegraphics[width=0.49\textwidth]{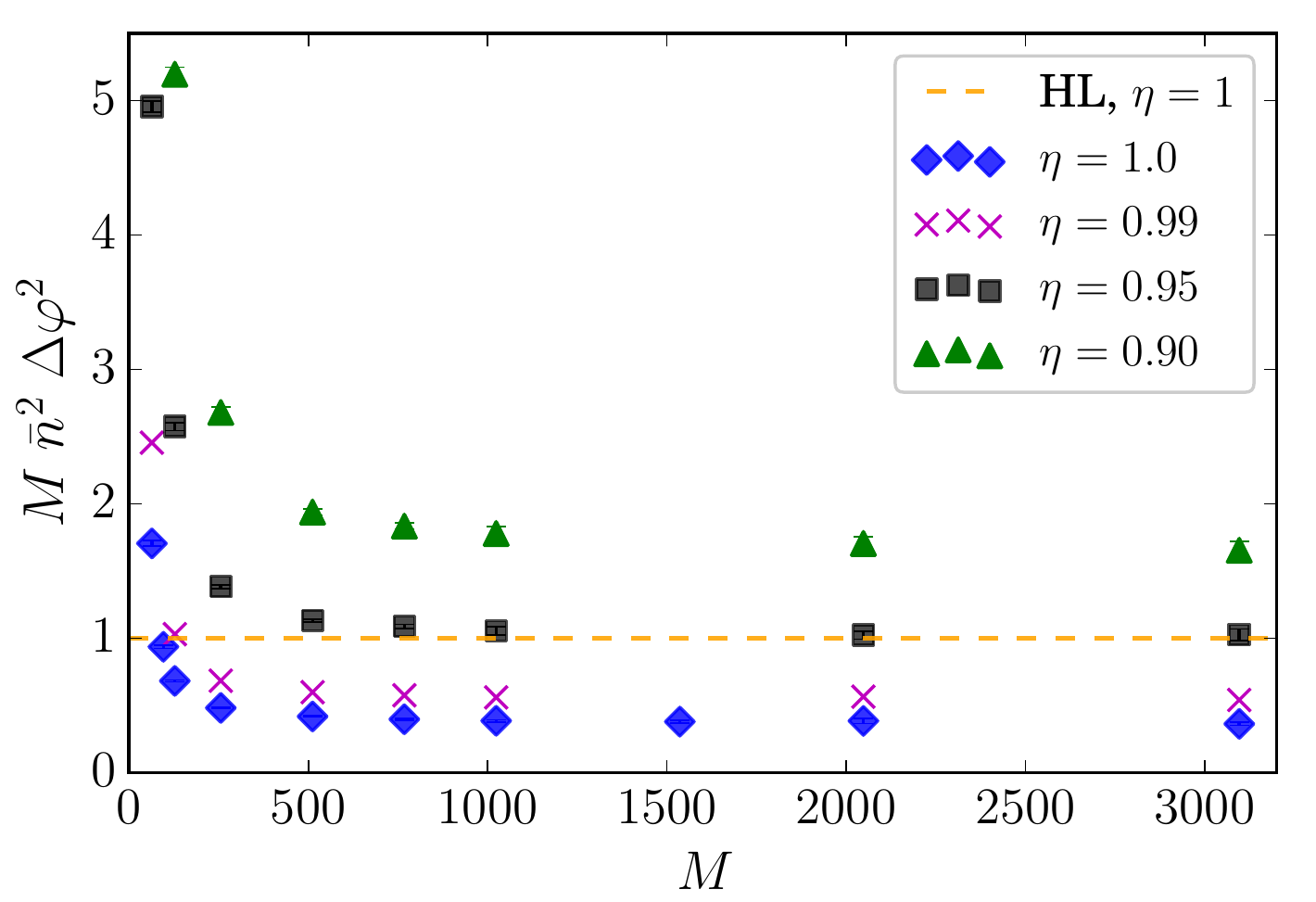} 
\caption{\label{fig:lossy_variance_1} 
The ratio of the phase MSE to the Heisenberg limit is plotted against the measurement record length $M$ for $\bar{n}=1$ and a range of levels of loss.
The results shown are: $\eta=1$ (blue diamonds), $\eta=0.99$ (magenta crosses), $\eta=0.95$ (gray squares), and $\eta=0.90$ (green triangles) are shown. The Heisenberg limit for $\eta = 1$ (dashed orange line) is plotted for comparison. The error bars are not shown if they are smaller than the markers.  } 
\end{figure}

\begin{figure}[tb]
\includegraphics[width=0.49\textwidth]{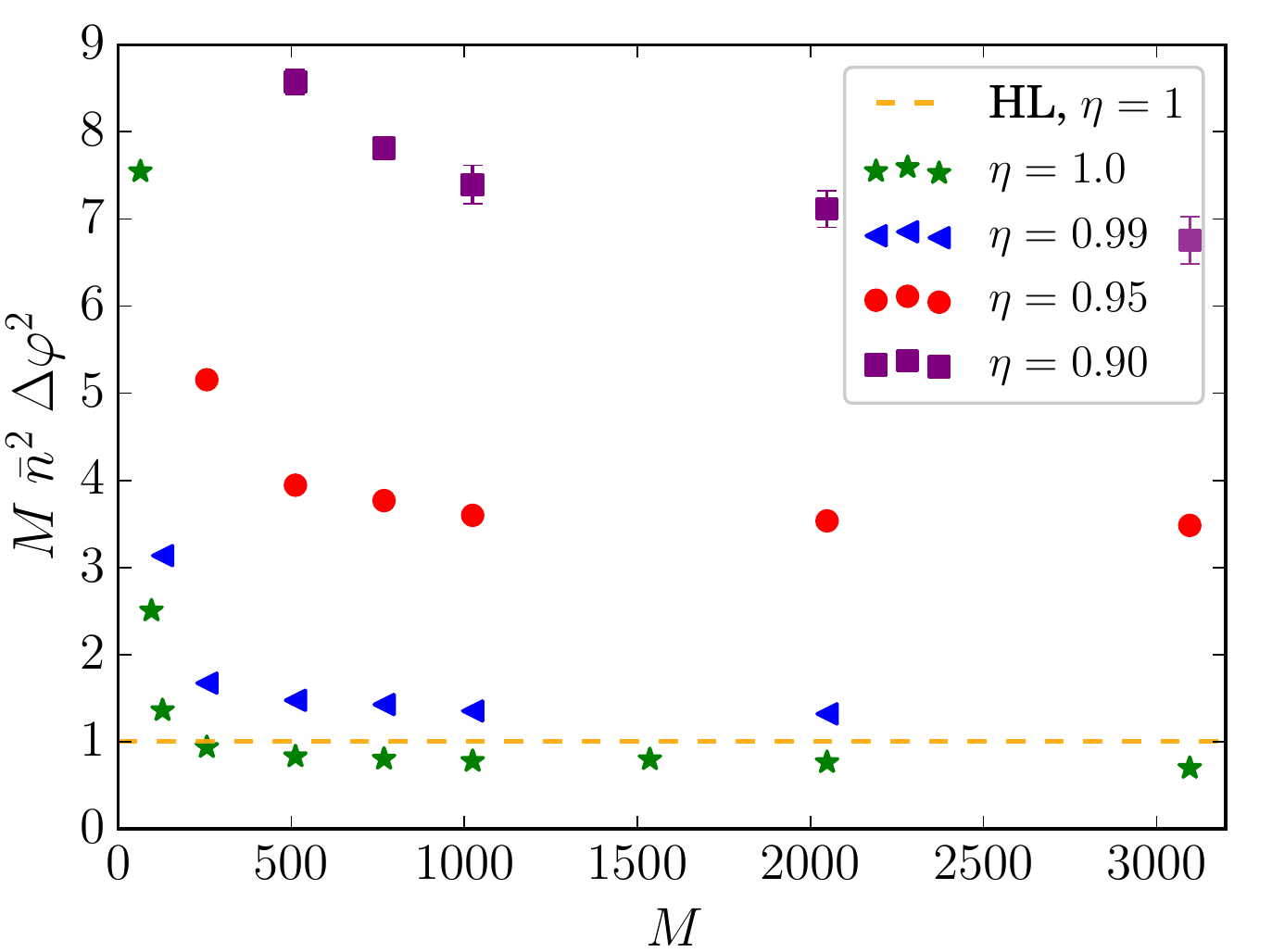} 
\caption{\label{fig:lossy_variance_3} 
The ratio of the phase MSE to the Heisenberg limit is plotted against the measurement record length $M$ for $\bar{n}=3$ and a range of levels of loss.
The results shown are: $\eta=1$ (green stars), $\eta = 0.99$ (blue triangles) and $\eta=0.95$ (red circles) and $\eta=0.9$ (purple squares) are shown. The Heisenberg limit for $\eta = 1$ (HL, dashed orange line) is plotted for comparison. The error bars are not shown if they are smaller than the markers.  } 
\end{figure}

Figures~\ref{fig:lossy_variance_1} and \ref{fig:lossy_variance_3} show the  phase MSE for $\bar{n}=1$ and $\bar{n}=3$ respectively. The MSE multiplied by $\bar{n}^2  M$ is plotted against $M$ for a range of values of $\eta$. The error bars are not shown if they are smaller than the marker size. For $\bar{n}=1$, when $\eta=0.95$, Heisenberg limit precision can be reached when $M $ is approximately one thousand, whereas for $\bar{n}=3$, the variance does not reach this limit even for $\eta=0.99$. As is evident in the plot, when losses are equal to $10\%$ for $\bar{n}=3$, the MSE increases by a factor of more than 10, whereas if single photon states or coherent states are used (mean-square error bounded by the shot-noise limit), the mean-square error would only increase by a factor of 1/$\eta$. Therefore, in order to observe error below the Heisenberg limit (or even the shot-noise limit), the system must be highly efficient.

\section{Conclusion}
\label{conc}

Schemes for phase measurement often consider N00N states, or equal photon numbers in both input ports for an interferometer \cite{Dowling2008,bib:holland1993interferometric,1367-2630-13-8-083026,giovannetti2011advances}.
The states that are most commonly produced experimentally are the TMSV states.
For these states, it was previously shown that parity detection yields a phase sensitivity, estimated from the Cram\'er-Rao bound, beyond the Heisenberg limit.
Since then, it was found that parity detection attains the Cram\'er-Rao bound for a wide range of states including TMSV states \cite{PhysRevA.87.043833}.

However, the ability to actually achieve estimation with mean-square error near the Cram\'er-Rao bound was uncertain, because the probability distribution for the measurements is not well behaved.
The sign of the phase is ambiguous, because for any nonzero result, the probability distribution for the phase is symmetric about zero.
Moreover, the phase measurement is most sensitive in a small region about zero, where the ambiguity of the sign is most problematic (because widely separated peaks would be easier to distinguish).

In this work we showed that it is possible to perform adaptive measurements on a moderate number of copies of the state, and achieve small mean-square error, that is close to the Cram\'er-Rao bound in the interval ($-\pi/2, \pi/2$). In particular, we find that it is possible to obtain mean-square error below the Heisenberg limit.
It was found that the higher the mean photon number is in the TMSV state, the larger the number of trials is needed to reach the Heisenberg limit. For mean photon number $\bar{n}=1$, approximately one hundred trials is sufficient, whereas for $\bar{n}=5$, one thousand is required. As the mean number of the TMSV state is increased, the number of parity detections needed to approximate the Cram\'er-Rao bound also increases. When loss is present, we find that the total efficiency of the system must be very close to one to observe precision below the Heisenberg limit: for mean photon number $\bar{n}= 3$, an efficiency larger than $0.99$ would be necessary.

\vspace{5mm}


\begin{acknowledgments}
DWB is funded by an ARC Future Fellowship (FT100100761) and an ARC Discovery Project (DP160102426). ZH is supported by the Australian Postgraduate Award scheme. KRM acknowledges the Australian Research Council Centre of Excellence for Engineered Quantum Systems (Project number CE110001013). JPD acknowledges support from the Air Force Office (Grant No.\ FA9550-13-10098), the Army Research Office (Grant No.\ W911NF-13-1-0381), the National Science Foundation (Grant No.\ 1403105), and the Northrop Grumman Corporation.
\end{acknowledgments}

%
\end{document}